\documentclass[aps,prl,twocolumn,groupedaddress,showpacs,floatfix]{revtex4}

\usepackage{graphicx}

\newcommand{\re}{{\bf r}  }
\newcommand{\rp}{{\bf r'}  }

\begin{document}


\title{Simple iterative construction of the optimized effective potential for
  orbital functionals, including exact exchange}

\author{Stephan K\"ummel}
\affiliation{Department of Physics and Quantum Theory Group, Tulane
  University, New Orleans, Louisiana 70118, USA}

\author{John P.\  Perdew}
\affiliation{Department of Physics and Quantum Theory Group, Tulane
  University, New Orleans, Louisiana 70118, USA}

\date{29 July 2002}

\begin{abstract}

For exchange-correlation functionals that depend explicitly on the
Kohn-Sham orbitals, the potential $V_{\mathrm{xc}\sigma}(\re)$ must be
obtained as the solution of the optimized effective potential (OEP)
integral equation. This is very demanding and has limited the use of
orbital functionals. We demonstrate that instead the OEP can be
obtained iteratively by solving the partial differential equations for
the orbital shifts that exactify the Krieger-Li-Iafrate (KLI)
approximation. Unoccupied orbitals do not need to be
calculated. Accuracy and efficiency of the method are shown for atoms
and clusters using the exact exchange energy. Counter-intuitive
asymptotic limits of the exact OEP are presented.

\end{abstract}


\maketitle

Density functional theory (DFT) has become one of the most successful
methods for electronic structure calculations. This success is largely
due to the availability of a "Jacob's ladder" \cite{jaclad} of
approximations to the exchange-correlation (xc) energy functional
$E_\mathrm{xc}$. High hopes for further improvements in functional
accuracy rest on third-generation functionals, i.e., functionals that
include full or partial exact exchange \cite{jaclad,hybrid1}: One more
exact contribution to the total energy can be taken into account, and
long-standing DFT problems like the self-interaction error and the
inexact high-density limit can be solved. Two problems so far stand in
the way of a widespread use of the exact exchange energy in DFT
calculations. Finding a correlation functional that is compatible with
exact exchange is the first; compatibility can be a severe problem for
the atomization energies of molecules. Hyper-generalized gradient
approximations \cite{jaclad} and
hybrid functionals \cite{hybrid1} are promising approaches to
solve this problem on either a non-empirical or empirical level. But
in order to use orbital-dependent functionals in self-consistent
Kohn-Sham calculations, a second problem -- constructing the
corresponding (spin-dependent) exchange-correlation potential
$V_\mathrm{xc\sigma}(\re)$ -- must be solved as well.  In fact,
solving the second problem might even help with the first, since known
constraints on the potential might be used to constrain the form of
$E_\mathrm{ c}$. But constructing $V_\mathrm{xc\sigma}(\re)$ for a
functional that depends explicitly on the Kohn-Sham orbitals (and thus
only implicitly on the density) is not straightforward: The potential
cannot be calculated directly as the functional derivative of the
exchange-correlation energy with respect to the density, but must be
obtained from the OEP integral equation \cite{sharp,talman,sahni}. A
direct solution of this integral equation is very demanding and so far
has only been achieved for effectively one-dimensional systems with
spherical symmetry \cite{talman,kli2,engel,grabo}. For solid state
systems, implementations within the atomic sphere approximation again
reduce the OEP problem to spherical symmetry \cite{kotani}.  For
systems without spherical symmetry, the OEP has recently been
constructed by directly evaluating the response function
\cite{goerlev,staedele,goer99,bartlett}. This makes molecular
calculations possible, but requires not only occupied but also
unoccupied orbitals and may lead to instabilities in the asymptotic
regions of the OEP, as discussed in \cite{hirata,lhf,hamel}, so
alternatives continue to be developed \cite{yang}. On the other hand,
the KLI \cite{kli2} and localized Hartree-Fock \cite{lhf,gritsenko}
approximations are easier to implement than the full OEP, yet accurate
in many cases \cite{kli2,lhf,gritsenko,grabo,rmmartinkli,engel2}.

We demonstrate how $V_\mathrm{xc\sigma}(\re)=\delta
E_\mathrm{xc}/ \delta n_\sigma(\re)$ can be calculated for any
orbital-dependent $E_\mathrm{ xc}$ without explicitly solving an
integral equation. The exact OEP is obtained iteratively by
self-consistently solving a system of partial differential equations
that is coupled to the Kohn-Sham equations. The iteration converges 
quickly. We use an essentially
asymptotically-correct approximation as the starting point, and the correct
asymptotic behavior is preserved during the iteration.
The unoccupied
Kohn-Sham orbitals need not be calculated. The orbital shifts we
evaluate were defined in Refs.\ \cite{kli2}, \cite{grabo}, but have
not been calculated before. However, orbital shifts have been used
successfully in linear response theory \cite{sternheimer,dfpt}, and
the advantages of avoiding explicit evaluation of the response function
are discussed 
in 
Ref.\ \cite{dfpt}.  We present self-consistent exact-exchange OEP
calculations for atoms and three-dimensional sodium clusters to
demonstrate the accuracy and efficiency of the method.

Excellent reviews of the OEP method have been published
\cite{grabo,engel2}, so we restrict ourselves to the central
equations. Following \cite{kli2,grabo}, the OEP integral equation is
written in the form 
\begin{equation}
\label{klisoep}
\sum_{i=1}^{N_\sigma} \psi_{i \sigma}^*(\re) \varphi_{i \sigma}(\re) +
c.c. 
=0.
\end{equation}
Here, c.c. denotes the complex conjugate of the previous term, and
$\varphi_{i \sigma}(\re)$ denotes Kohn-Sham orbitals, i.e., the solutions of
\begin{equation}
\label{kseq}
\left(\hat{h}_{\mathrm{KS}\sigma}-\varepsilon_{i \sigma}\right)
 \varphi_{i \sigma}(\re)= 0,
\end{equation}
where 
$\hat{h}_{\mathrm{KS}\sigma}=-(\hbar^2/2m) \nabla^2
 +V_{\sigma}(\re)$ is the Kohn-Sham Hamiltonian,  
$V_{\sigma}(\re)=V_\mathrm{ext}(\re)+V_\mathrm{H}(\re)+
V_{\mathrm{xc}\sigma}(\re)$ is the Kohn-Sham potential with
$V_\mathrm{ext}(\re)$ the external potential and
$V_\mathrm{H}(\re)=\int d^3 r' e^2 n(\rp)/|\re -\rp| $ the Hartree potential.
The unnormalized orbital shifts
$\psi_{i \sigma}^*(\re)$ are defined by
\begin{eqnarray}
\label{psipert}
\lefteqn{
\psi_{i \sigma}^*(\re)=}
\\ & &
\sum_{\stackrel{\scriptstyle j=1}{j\ne i}}^{\infty}
\frac{\int \varphi_{i \sigma}^*(\rp)
  \left[ V_{\mathrm{xc}\sigma}(\rp)-u_{\mathrm{xc}i\sigma}(\rp) \right]
  \varphi_{j \sigma}(\rp) \, \mathrm{d}^3 r'}
{\varepsilon_{i \sigma}-\varepsilon_{j \sigma} }
\varphi_{j \sigma}^*(\re), \nonumber 
\end{eqnarray}
where
\begin{equation}
u_{\mathrm{xc}i\sigma}(\re)=\frac{1}{\varphi_{i \sigma}^*(\re)}
 \frac{\delta E_\mathrm{xc}\left[\left\{ \varphi_{j \tau}
                                 \right\}
                          \right]}
      {\delta \varphi_{i \sigma}(\re)}.
\end{equation}
(Ref. \cite{kli2} uses 
$p_{i\sigma}(\re)=-\psi_{i \sigma}^*(\re)/\varphi^*_{i\sigma}$.)
The $-\psi_{i \sigma}(\re)$ are the
first-order perturbation theory corrections to the
$\varphi_{i\sigma}(\re)$ for the perturbing potential
$u_{\mathrm{xc}i\sigma}(\re)-V_{\mathrm{xc}\sigma}(\re)$. 
Thus, Eq.\ (\ref{klisoep}) asserts that the density shift
vanishes to first order, and 
perturbation theory shows that 
$\psi_{i \sigma}^*(\re)$ must fulfill the partial differential equation
\begin{eqnarray}
\label{psipde}
\lefteqn{
(\hat{h}_{\mathrm{KS}\sigma}-\varepsilon_{i \sigma})\psi_{i \sigma}^*(\re)=
} \nonumber \\ & &
-[V_{\mathrm{xc}\sigma}(\re)-u_{\mathrm{xc}i\sigma}(\re)-
   (\bar{V}_{\mathrm{xc}i\sigma}-\bar{u}_{\mathrm{xc}i\sigma})
 ]\varphi_{i\sigma}^*(\re).
\end{eqnarray}
In Eq.\ (\ref{psipde}), the orbital averages are
$
\bar{V}_{\mathrm{xc}i\sigma}=\int \varphi_{i\sigma}^*(\re) 
  V_{\mathrm{xc}\sigma}(\re)\varphi_{i\sigma}(\re) \, \mathrm{d}^3r
$
and
$
\bar{u}_{\mathrm{xc}i\sigma}=\int \varphi_{i\sigma}^*(\re) 
 u_{\mathrm{xc}i\sigma}(\re) \varphi_{i\sigma}(\re) \, \mathrm{d}^3r
$.
Multiplying Eq.\ (\ref{klisoep}) by $V_{\sigma}(\re)$ and inserting
Eq.\ (\ref{psipde}), solved for $V_{\sigma}(\re)$, into the resulting
expression, an equation is obtained that can be solved for
$V_{\mathrm{xc}\sigma}(\re)$, and a few further 
manipulations \cite{kli2,grabo} then lead to the form
\begin{eqnarray}
\label{oepexpl}
\lefteqn{
V_{\mathrm{xc}\sigma}(\re)=}
\nonumber \\ & &
\frac{1}{2 n_\sigma(\re)}
\sum_{i=1}^{N_\sigma}\left\{ 
\left| \varphi_{i\sigma}(\re) \right|^2 \left[ 
  u_{\mathrm{xc}i\sigma}(\re)+ 
  (\bar{V}_{\mathrm{xc}i\sigma}-\bar{u}_{\mathrm{xc}i\sigma}) \right]
\right.  \nonumber \\ & & \left. 
-\frac{\hbar^2}{m}\nabla \cdot \left[\psi_{i \sigma}^*(\re)\nabla 
        \varphi_{i\sigma}(\re) \right] \right\} + c.c.
\end{eqnarray}
The KLI potential is obtained by setting $\psi_{i\sigma}^*(\re)=0 \, \forall
\, i$ in the above equation, a choice that can be motivated in the
sense of a mean-field approximation \cite{kli2,grabo}.

Combining Eqs.\ (\ref{kseq}) and (\ref{psipde}) to calculate
$V_{\mathrm{xc}\sigma}(\re)$ in a self-consistent iteration is the
basic idea of our method. Calculating the $\psi_{i\sigma}^*(\re)$ is thus the
first crucial step.
The first problem to be
addressed is which boundary conditions are appropriate. Since the
$\psi_{i\sigma}^*(\re)$ are orbital shifts, the boundary condition to be 
employed in the solution of Eq.\ (\ref{psipde}) must be the same as
the one used in solving Eq.\ (\ref{kseq}), i.e, for our finite
systems, $\psi_{i \sigma}^*(\re)\rightarrow0$ for 
$r\rightarrow\infty$ \cite{grabo}. But at first sight, calculating
$\psi_{i\sigma}^*(\re)$ still seems a hopeless task: Evaluating Eq.\
(\ref{psipert}) is highly impractical and Eq.\ (\ref{psipde}) is singular. It
does not have a unique 
solution because to any one solution $\psi_{i\sigma}(\re)$, the Kohn-Sham
orbital $\varphi_{i\sigma}(\re)$ multiplied by an arbitrary constant can be
added to find another solution.

That Eq.\ (\ref{psipde}) can nevertheless be solved efficiently is
based on two facts. First, it is clear from Eq.\ (\ref{psipert}) that
the particular solution of Eq.\ (\ref{psipde}) that is relevant for
the OEP must be orthogonal to the Kohn-Sham orbital $\varphi_{i
\sigma}(\re)$. Second, the right hand side of Eq.\ (\ref{psipde}) is
orthogonal to $\varphi_{i \sigma}(\re)$.  The orbital shift therefore
can be calculated elegantly by the conjugate gradient method
\cite{numrec}. In short, if the operator on the left hand side of Eq.\
(\ref{psipde}) is abbreviated by $\hat{A}_i$ and the right hand side
by $b_i$, then the conjugate gradient method constructs the solution
of $\hat{A}_i\psi_i^*=b_i$ by calculating a sequence of additive
corrections to a starting guess $\psi_{i\mathrm{s}}^*$. The first
correction in the sequence is, up to a constant factor, given by the
expression $b_i-\hat{A}_i\psi_{i\mathrm{s}}^*$. Since $b_i$ is
orthogonal to $\varphi_i$, and since $\varphi_i$ spans the nullspace
of $\hat{A}_i$, the whole correction is orthogonal to
$\varphi_i$. Since the following steps in the sequence proceed
analogously \cite{numrec}, the desired orthogonal solution is obtained
if $\psi_{i\mathrm{s}}^*$ is chosen orthogonal. In practice, numerical
inaccuracies might lead to small contributions of $\varphi_i$ to the
calculated $\psi_i$, but they can easily be removed by orthogonalizing
the final $\psi_i$ to $\varphi_i$ (or, in the case of degeneracies, to
all Kohn-Sham orbitals having the eigenvalue $\varepsilon_i$). Real
space and basis-set implementations should be equally unproblematic
since $V_\mathrm{x}(r \rightarrow \infty)$ is essentially correct in
the starting approximation.

One way of constructing the OEP would be to insert the orbital shifts
into Eq.\ (\ref{oepexpl}). But this is numerically cumbersome for
finite systems: Evaluating the last term on the right hand side of
Eq.\ (\ref{oepexpl}) requires dividing derivatives of exponentially
decaying functions by the exponentially decaying density. For large
distances where all orbitals have decayed to the extent that their
numerical representation comes close to zero, inaccuracies can be
introduced into the iteration. We found that, with appropriate
cut-offs, the iteration converged quickly for the Be atom. However,
for non-spherical systems this procedure becomes cumbersome.

\begin{figure*}[bht]
\includegraphics[width=8.4cm]{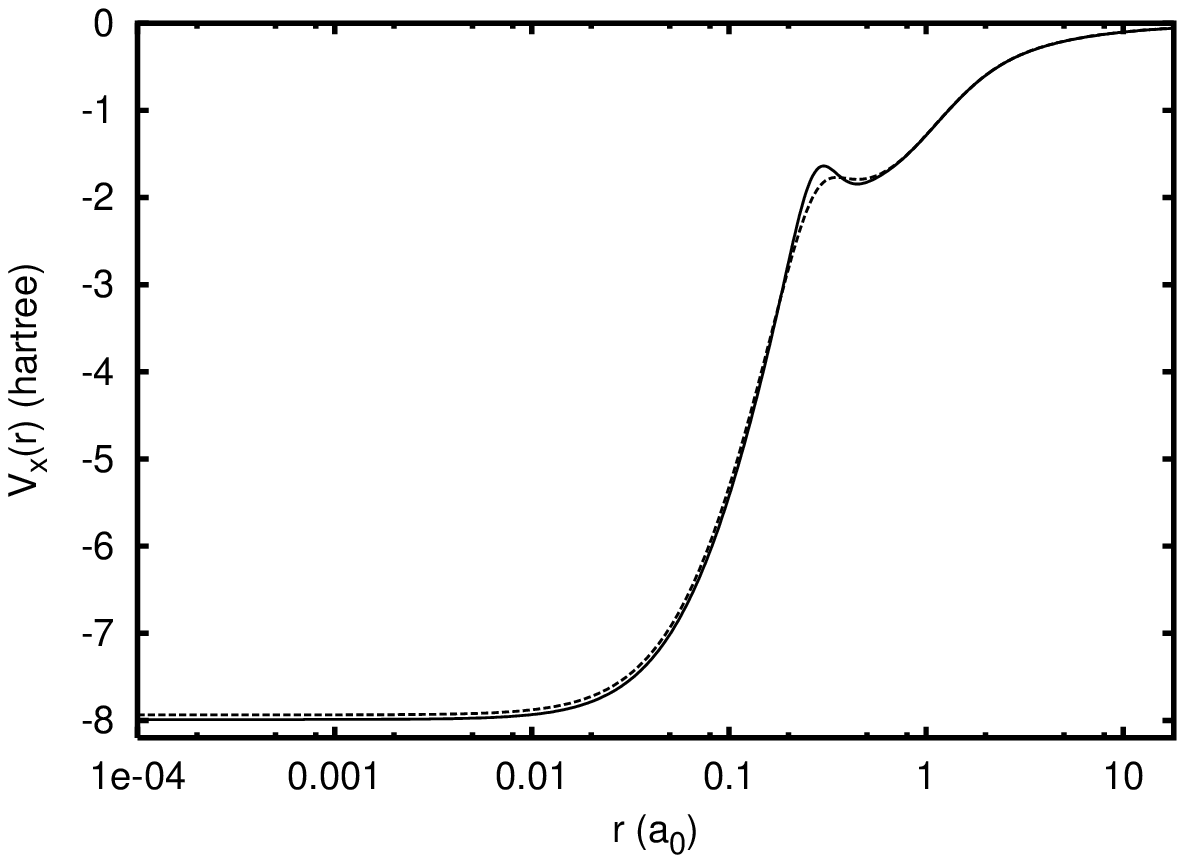}
\includegraphics[width=8.4cm]{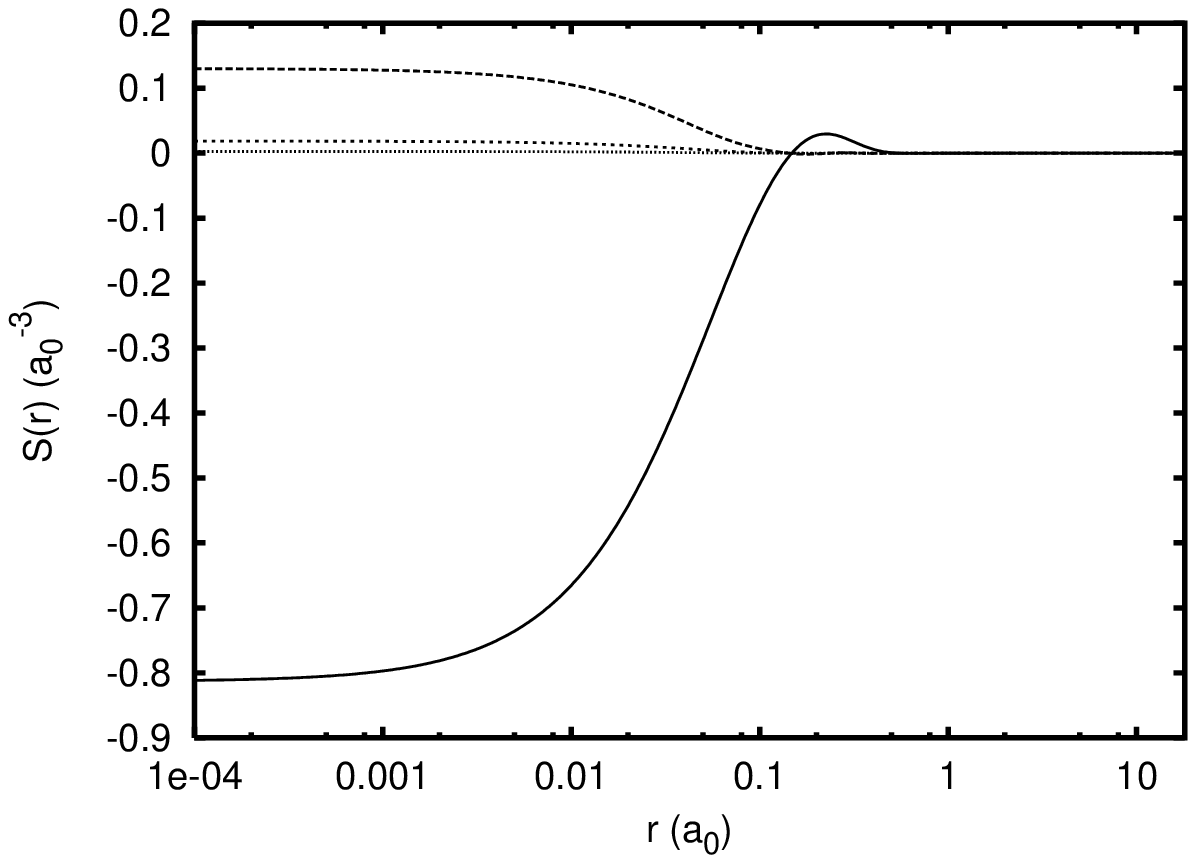}
\caption{Left: $V_\mathrm{x}(\re)$ for the Ne
atom versus radial coordinate $r$ in bohr $(a_0)$. Dashed: KLI. Full: OEP.
Right: $S(r)$
at different stages of the iteration. 
Full: $S(r)$ for KLI potential. Long dashed, short dashed
and dotted: $S(r)$ after first, second and third iteration. Note that
the magnitude of $S(r)$ is reduced by each iteration and goes to zero.
\label{fig1}} 
\end{figure*}
An alternative way of using the orbital shifts to construct the OEP is
based on the fact that once the $\psi_{i \sigma}^*(\re)$ have been
calculated for a given 
$V_\mathrm{xc\sigma}(\re)$, one obtains from them an estimate for
how close the given $V_\mathrm{xc\sigma}(\re)$ is to the true OEP in
the region of nonvanishing density by defining the function
\begin{equation}
\label{defs}
S_\sigma(\re)= 
\sum_{i=1}^{N_\sigma} \psi_{i \sigma}^*(\re) \varphi_{i \sigma}(\re) 
+ c.c.
\end{equation}
For the exact OEP, Eq.\ (\ref{defs}) reduces to Eq.\ (\ref{klisoep}),
i.e., $S_\sigma(\re)=0$ $\forall \, \re$. But if
$\varphi_{i\sigma}(\re)$ and $\psi_{i \sigma}^*(\re)$ have been
calculated via Eqs.\ (\ref{kseq}) and (\ref{psipde}) for an
approximate $V_\mathrm{xc\sigma}(\re)$, then $S_\sigma(\re)$ does not
vanish. As an example, the left half of Fig.\ \ref{fig1} shows the OEP
and KLI potentials for the Ne atom, and the full line in the right
half of Fig.\ \ref{fig1} shows the $S(r)$ obtained for the KLI
potential. (When there is no spin polarization, we drop the index
$\sigma$.) Although $S(r)$ is not an exact representation of the
difference between the two potentials, as can be seen by comparing
$S(r)$ to Fig.\ 2 of Ref.\ \cite{kli2}, it nevertheless indicates
where and qualitatively how the approximation deviates from the OEP:
$S(r)$ vanishes in the asymptotic region where KLI is correct, but
features the greater intershell bump and lower core value of the OEP.

This observation motivates the following simple iterative construction
of the OEP: Using an approximation to the exact OEP, calculate 
$\varphi_{i\sigma}(\re)$ and $\psi_{i \sigma}^*(\re)$ via
Eqs.\ (\ref{kseq}) and (\ref{psipde}) and $S_\sigma(\re)$ via Eq.\
(\ref{defs}). An improved  approximation to the OEP is
then constructed as
\begin{equation}
\label{iter}
V_\mathrm{xc\sigma}^\mathrm{new}(\re) =
V_\mathrm{xc\sigma}^\mathrm{old}(\re) + c S_\sigma(\re),
\end{equation}
where $c$ is a positive constant having the dimension of energy over
density. The idea of this iteration is to remove the error -- for
which $S_\sigma(\re)$ is a measure -- from the current approximation
to obtain a better one. The numerical value of $c$ is a
system-dependent parameter of the iteration. Larger values lead to
faster convergence, but too large a value will ultimately lead to
divergence. $c$ is thus similar to the parameter introduced in the
overrelaxation method \cite{numrec} or the mixing parameter in
self-consistent calculations. In the calculations discussed below we
used values of about 1 for the atoms and of about 30 for the clusters
on a trial-and-error basis. A rationale for these values will be
presented in future work. In a simple but important second step, it
must be ensured that the improved approximation satisfies the relation
$\bar{V}_{\mathrm{xc}N_\sigma\sigma}=\bar{u}_{\mathrm{xc}N_\sigma\sigma}$ 
which is obeyed by the exact OEP \cite{kli2,grabo}. This can easily be
achieved by adding the constant
$\bar{u}_{\mathrm{xc}N_\sigma\sigma}-\int \varphi_{N_\sigma\sigma}^*(\re) 
  V_{\mathrm{xc}\sigma}^\mathrm{new}(\re)\varphi_{N_\sigma\sigma}(\re)
\, \mathrm{d}^3r
$
to the $V_{\mathrm{xc}\sigma}^\mathrm{new}(\re)$ computed from Eq.\
(\ref{iter}). With this new approximation for
$V_{\mathrm{xc}\sigma}(\re)$, Eqs.\ (\ref{kseq}) and (\ref{psipde})
can be solved anew to start another cycle until convergence is achieved.

In our calculations the most efficient way of doing one iteration of Eq.\
(\ref{kseq}) is to do several cycles of iterating $\{\psi_{i\sigma}^*(\re)\}$
and $V_{\mathrm{xc}\sigma}(\re)$ while keeping $\{\varphi_{i \sigma}(\re)\}$,
$\{u_{\mathrm{xc}i\sigma}(\re)\}$, $\hat{h}_\mathrm{KS}$ and $\varepsilon_i$
in Eq.\ (\ref{psipde}) fixed. Only then are the Kohn-Sham equations solved
anew. The reason for this is that in our real-space finite difference
calculations, solving Eq.\ (\ref{psipde}) requires only roughly as much
computational effort as calculating the Hartree potential, whereas solving the
Kohn-Sham eigenvalue equations is more demanding.

\begin{table}[b]
\caption{Total energy and Kohn-Sham eigenvalues in hartree
for the Ne (top) and Ar (bottom) atom. First line (iter=0): KLI
approximation. Second line: OEP. The first column gives the
iteration number at which all eigenvalues are converged to $0.0001$
hartree accuracy. Total energies are already correct after the first
iteration.}  
\label{tab1}
\begin{ruledtabular}
\begin{tabular}{ccccccc}
iter& E & $\varepsilon_1$ &$\varepsilon_2$ &$\varepsilon_3$ &$\varepsilon_4$ 
&$\varepsilon_5$  \\ \hline
0   & -128.5448 & -30.8021 & -1.7073  & -0.8494 & & \\
3   & -128.5454 & -30.8200 & -1.7181  & -0.8507 & &  \\ \hline
0   & -526.8105 & -114.4279 & -11.1820 & -8.7911 & -1.0942 & -0.5893  \\
6   & -526.8122 & -114.4522 & -11.1532 & -8.7338 & -1.0993 & -0.5908  \\
\end{tabular}
\end{ruledtabular}
\end{table}
We first employed this scheme to calculate OEP total energies and
eigenvalues of spherical atoms in the exchange-only
approximation. The corresponding equations were solved on a
logarithmic grid. The right half of Fig.\
\ref{fig1} shows that each iteration reduces $S$, as it
should. The first line of Table \ref{tab1} shows the KLI
energies, the starting point of our iteration. The
second line gives the same values for the OEP, and the
first column indicates how many iterations were necessary to converge
to 0.0001 hartree accuracy. Our simple iteration converges
within a few steps to very high accuracy. We have carefully
checked that these numbers correspond to the ones obtained by the much
more involved direct solution of the OEP integral equation, by
re-running the program used in Refs.\ \cite{engel,grabo} on
fine grids. 

Our scheme can easily be used to calculate the
OEP for three-dimensional systems without particular symmetry. To
demonstrate this, we have calculated the OEP for the 
sodium clusters $\mathrm{Na}_4$ and $\mathrm{Na}_8$, using the
pseudopotential and geometries from Ref.\ \cite{napp}. A
three-dimensional Cartesian grid with 
129 points in each direction and of spacing 0.5$a_0$ was employed to obtain an
accuracy of 0.0001 hartree. Eq.\ (\ref{kseq}) was solved by damped
gradient iteration with multigrid relaxation, and 
Eq.\ (\ref{psipde}) by our own conjugate gradient
routine. 
In Table \ref{tab2} we list total energies and eigenvalues obtained
with the exact exchange functional, iterating from KLI to OEP. 
The first thing to note is that, even for the clusters, KLI is a
rather good approximation to the OEP. But it should also be noted that
the relative error in the KLI energies is nearly two 
orders of magnitude higher for the clusters than for the
atoms. Nevertheless, just five steps of our iteration refine KLI 
to OEP energies. 
The relative error in the
Levy-Perdew virial relation \cite{virial} falls to $\sim 10^{-5}$
during the first five iterations, and to $\sim 10^{-7}$ upon
further iteration, demonstrating the accuracy and stability of our
method. 
\begin{table}[t]
\caption{KLI and OEP total energies and eigenvalues
from three-dimensional cluster calculations, listed as in Table
\ref{tab1}. Top: $\mathrm{Na}_4$, Bottom: $\mathrm{Na}_8$. See text
for details.}    
\label{tab2}
\begin{ruledtabular}
\begin{tabular}{cccccc}
iter& E & $\varepsilon_1$ &$\varepsilon_2$ &$\varepsilon_3$ &$\varepsilon_4$ 
\\ \hline
0   & -0.7528 & -0.1776  & -0.1394 & & \\
5   & -0.7531 & -0.1762 &  -0.1401 & & \\ \hline
0   & -1.5282 & -0.1963 & -0.1591 & -0.1458 & -0.1458  \\
5   & -1.5285 & -0.1954 & -0.1592 & -0.1459 & -0.1459  \\
\end{tabular}
\end{ruledtabular}
\end{table}

Finally, we investigated the asymptotic behavior of the Kohn-Sham
potential. Contrary to common belief, the exchange potential does not
fall off asymptotically like $-e^2/r$ everywhere for finite systems,
but approaches $-e^2/r + C $, where $C$ is a nonvanishing constant, on
nodal surfaces of the energetically-highest occupied orbital. For
$\mathrm{Na}_4$ we find $C=0.0307$ for OEP and $C=0.0298$ hartree for
KLI. This unexpected behavior of $V_\mathrm{x}(\re)$ can be understood
on the basis of Eq.\ (\ref{oepexpl}). It is not reproduced by any of
the common methods such as the local density approximation, and also
not by the OEP construction presented in Ref.\ \cite{yang}. Our work
thus demonstrates the existence of non-vanishing asymptotic
constants in the exact OEP, and confirms the arguments recently
put forward in Ref.\ \cite{asymp}.

In conclusion, our simple iterative construction of
the OEP does not require solving an integral equation or
calculating unoccupied Kohn-Sham orbitals. We discussed the surprising
asymptotic behavior of $V_\mathrm{x}(\re)$. All-electron calculations
for atoms and three-dimensional pseudopotential calculations for
clusters demonstrate the accuracy and broad range of applicability of
our approach. It thus opens a path for employing orbital
functionals, in particular the exact exchange energy, in
self-consistent Kohn-Sham calculations.

\begin{acknowledgments}
S.K.\ acknowledges financial support by the Deutsche
Forschungsgemeinschaft under an Emmy-Noether grant and J.P.P.\ by the
U.S.\  National Science Foundation under grant DMR 01-35678. 
\end{acknowledgments}





\begin{thebibliography}{100}


\bibitem{jaclad}J. P. Perdew and K. Schmidt,
in {\it Density Functional Theory and Its Applications to Materials},
edited by V. VanDoren et al.
(American Institute of Physics, 2001).

\bibitem{hybrid1}A. D. Becke, J. Chem. Phys. {\bf 107}, 8554 (1997).

\bibitem{sharp}R. T. Sharp and G. K. Horton, Phys. Rev. {\bf 90}, 317 (1953).

\bibitem{talman}J. D. Talman and W. F. Shadwick, Phys. Rev. A 
  {\bf 14}, 36 (1976). 

\bibitem{sahni}V. Sahni, J. Gruenebaum, and J. P. Perdew, Phys. Rev. B
{\bf 26}, 4371 (1982).

\bibitem{kli2}J. B. Krieger, Y. Li, and G. J. Iafrate,
Phys. Rev. A {\bf 46}, 5453 (1992).

\bibitem{engel}E. Engel and S. H. Vosko, Phys. Rev. A {\bf 47}, 2800
(1993); Phys. Rev. B {\bf 50}, 10498 (1994).

\bibitem{grabo}T. Grabo {\it et al.}, 
in {\it Strong Coulomb Correlation in Electronic Structure},
edited by V. I. Anisimov (Gordon \& Breach, Tokyo, 2000), pp. 203-311;
Molecular Engineering {\bf 7}, 27 (1997).

\bibitem{kotani}
T.\ Kotani and H.\ Akai, Phys.\ Rev.\ B {\bf 54}, 16502 (1996).

\bibitem{goerlev}A. G\"orling and M. Levy, Phys. Rev. A {\bf 50}, 196 (1994).

\bibitem{staedele}M. St\"adele {\it et al.},
Phys. Rev. Lett. {\bf 79}, 2089 (1997); 
Phys. Rev. B {\bf 59}, 10031 (1999).

\bibitem{goer99}A. G\"orling, Phys. Rev. Lett. {\bf 83}, 5459 (1999).

\bibitem{bartlett}S. Ivanov, S. Hirata, and R. J. Bartlett,
Phys. Rev. Lett. {\bf 83}, 5455 (1999).

\bibitem{hirata}S. Hirata {\it et al.},
J. Chem. Phys. {\bf 115}, 1635 (2001).

\bibitem{lhf}F. Della Sala and A. G\"orling, J. Chem. Phys. {\bf
115}, 5718 (2001).

\bibitem{hamel}S.\ Hamel, M.\ E.\ Casida, and D.\ R.\ Salahub, J.\
Chem.\ Phys.\ {\bf 116}, 8276 (2002).

\bibitem{yang}
W.\ Yang and Q.\ Wu, Phys.\ Rev.\ Lett.\ {\bf 89}, 143002 (2002). 

\bibitem{gritsenko}O.\ V.\ Gritsenko and E.\ J.\ Baerends, Phys.\
Rev.\ A {\bf 64}, 042506 (2001); S. J. A. van Gisbergen {\it et al.},
Phys. Rev. Lett. {\bf 83}, 694 (1999). 

\bibitem{rmmartinkli}Y.-H. Kim, M. St\"adele, and R.M. Martin, 
Phys. Rev. A {\bf 60}, 3633 (1999).

\bibitem{engel2}E. Engel and R. M. Dreizler, J. Comp. Chem. {\bf 20},
31 (1999).

\bibitem{sternheimer}R.\ M.\ Sternheimer, Phys.\ Rev.\ {\bf 96}, 951 (1954).

\bibitem{dfpt}S.\ Baroni {\it et al.}, 
Phys.\ Rev.\ Lett.\ {\bf 58}, 1861 (1987); Rev.\ Mod.\ Phys.\ {\bf
73}, 515 (2001).

\bibitem{numrec}See, e.g., W.\ H.\ Press {\it et al.}, 
{\it Numerical Recipes in FORTRAN}  (Cambridge University Press, 1992).

\bibitem{napp}S.\ K\"ummel, M.\ Brack, and P.-G.\ Reinhard,
Phys.\ Rev.\ B \textbf{62}, 7602 (2000); ibid. {\bf 63},
129902 (2001) (E). 

\bibitem{virial}M.\ Levy and J.\ P.\ Perdew, Phys.\ Rev.\ A {\bf 32}, 2010 (1985).

\bibitem{asymp}F. Della Sala and A. G\"orling, Phys. Rev. Lett. {\bf
89}, 33003 (2002).   

\end{thebibliography}
\end{document}